%% file: ms.tex
\documentclass[12pt,article]{emulateapj} 
\shorttitle{Direct Evidence of Cold Gas in \dla 0812$ +$32B}
\shortauthors{Jorgenson et al.}

\input{xavier_defs.tex}

\begin{document}

\def\intl{\int\limits}
\def\nstat{$\approx $}
\def\perd{\;\;\; .}
\def\cmma{\;\;\; ,}
\def\ltk{\left [ \,}
\def\ltp{\left ( \,}
\def\ltb{\left \{ \,}
\def\rtk{\, \right  ] }
\def\rtp{\, \right  ) }
\def\rtb{\, \right \} }
\def\jnu{$J_{\nu}$}
\def\jnuphot{$J_{\nu}^{phot}$}
\def\jnuciistar{$J_{\nu}^{\rm CII^{*}}$}
\def\junit{ergs cm$^{-2}$ s$^{-1}$ Hz$^{-1}$ sr$^{-1}$}
\def\jnutot{$J_{\nu}$$^{total}$}
\def\jnubkd{$J_{\nu}$$^{Bkd}$}
\def\jnuloc{$J_{\nu}$$^{local}$}
\def\jnulw{$J_{\nu}$$^{LW}$}
\def\jnutotciistr{$J_{\nu }$$^{total, C\,II^*}$}
\def\jnulocciistr{$J_{\nu } $$^{local, C\,II^*}$}
\def\jnutotci{$J_{\nu }$$^{total, C\, I}$}
\def\jnulocci{$J_{\nu }$$^{local, C\, I}$}
\def\jnutothtwo{$J_{\nu }$$^{total, H_2}$}
\def\jnulochtwo{$J_{\nu }$$^{local, H_2}$}
\newcommand{\snrlim}{S/N$_{lim}$}
\newcommand{\nhi}{$N_{\rm HI}$}
\newcommand{\mnhi}{N_{\rm HI}}
\newcommand{\flls}{f_{\rm HI}^{\rm LLS}}
\newcommand{\fdla}{f_{\rm HI}^{\rm DLA}}
\newcommand{\llls}{$\ell_{\rm LLS}$}
\newcommand{\ldla}{\ell_{\rm DLA}}
\newcommand{\fnhi}{$f_{\rm HI}(N,X)$}
\newcommand{\mfnhi}{f_{\rm HI}(N,X)}
\newcommand{\Nth}{2 \sci{20} \cm{-2}}
\newcommand{\taux}{$d\tau/dX$}
\newcommand{\gz}{$g(z)$}
\newcommand{\nz}{$n(z)$}
\newcommand{\nx}{$n(X)$}
\newcommand{\omg}{$\Omega_g$}
\newcommand{\ostr}{$\Omega_*$}
\newcommand{\momg}{\Omega_g}
\newcommand{\olls}{$\Omega_g^{\rm LLS}$}
\newcommand{\odla}{$\Omega_g^{\rm DLA}$}
\newcommand{\oneut}{$\Omega_g^{\rm Neut}$}	
\newcommand{\ohi}{$\Omega_g^{\rm HI}$}
\newcommand{\olwz}{$\Omega_g^{\rm 21cm}$}
\newcommand{\ndla}{71}
\newcommand{\cmk}{cm$^{-3}$ K }
\newcommand{\nv}{N\,V}
\newcommand{\ovi}{O\,VI}
\newcommand{\ci}{C\,I}
\newcommand{\cistr}{C\,I$^{*}$}
\newcommand{\mcistr}{C\,I^{*}}
\newcommand{\cistrstr}{C\,I$^{**}$}
\newcommand{\mcistrstr}{C\,I^{**}}
\newcommand{\citot}{(C\,I)$_{tot}$}
\newcommand{\mcitot}{(C\,I)_{tot}}
\newcommand{\cli}{Cl\,I}
\newcommand{\clii}{Cl\,II}
\newcommand{\cii}{C\,II}
\newcommand{\ciistr}{C\,II$^*$}
\newcommand{\dla}{DLA}
\newcommand{\dlas}{DLAs}
\newcommand{\htwo}{H$_{\rm 2}$}
\newcommand{\he}{He\, I}
\newcommand{\sii}{Si\,II}
\newcommand{\siistr}{Si\,II$^{*}$}
\newcommand{\hi}{H\, I}
\newcommand{\ctwo}{\ion{C}{2}}
\def\lc{${\ell}_{c}$}
\def\lcunit{ergs s$^{-1}$ H$^{-1}$}

\title{Direct Evidence of Cold Gas in \dla\ 0812$ +$32B}

\author{Regina A. Jorgenson\altaffilmark{1}, Arthur M. Wolfe\altaffilmark{2}, J. Xavier Prochaska\altaffilmark{3}, Robert F. Carswell\altaffilmark{1}, 
}

\altaffiltext{1}{Institute of Astronomy, University of Cambridge, Madingley Road, Cambridge, CB3 0HA, UK; raj@ast.cam.ac.uk}

\altaffiltext{2}{Department of Physics, and 
Center for Astrophysics and Space Sciences, 
University of California, San Diego, 
9500 Gilman Dr., La Jolla; CA 92093-0424}

\altaffiltext{3}{Department of Astronomy and Astrophysics, 
UCO/Lick Observatory;
University of California, 1156 High Street, Santa Cruz, CA  95064}

\begin{abstract}
We present the first direct evidence for cold gas in a high redshift \dla\ galaxy.  We measured several multiplets of weak neutral carbon (\ci\ ) transitions in order to perform a curve of growth analysis.  A $\Delta$$\chi$$^2$ test constrains the best fit Doppler parameter, 
$b = 0.33 _{-0.04}^{+0.05}$\kms , and logN(\ci ) = $13.30 \pm 0.2$ cm$^{-2}$.  
This Doppler parameter constrains the kinetic temperature of the gas to T $\leq$ 78 K (T $\leq$ 115 K, 2 $\sigma$).  We used the associated \ci\ fine structure lines to constrain the volume density of the gas, n(\hi ) $\sim 40 - 200$\,cm$^{-3}$ (2$\sigma$), resulting in a lower limit on the cloud size of approximately 0.1 $-$ 1 parsec.  While it is difficult to determine the metallicity of the cold component, the absence of Cr II indicates that the cold cloud suffers a high level of dust depletion.  Additionally, the large amount of Lyman and Werner-band molecular hydrogen absorption (log N(\htwo )$_{total}$ = 19.88 cm$^{-2}$, f$_{H_2} $ $\geq$ 0.06) with an excitation temperature of T$_{ex}$ = 46 K as determined by the rotational J = 0 and J = 1 states, is consistent with the presence of cold gas.  We propose that this cloud may be gravitationally confined and  may represent a transition gas-phase from primarily neutral atomic gas, to a colder, denser molecular phase that will eventually host star formation.

\end{abstract}


\section{Introduction}

As high redshift neutral gas reservoirs for star formation, Damped Lyman alpha Systems (\dlas ) are key components for understanding galaxy formation and evolution.  \cite{wolfe03a} suggested
that the gas in \dlas\ exists in a multiphase medium, with cold and warm components in pressure equilibrium.  This model is important in explaining the pressure confinement of the gas, as well as in determining the physical properties of the gas such as temperature and density.  However, a direct measurement of the temperature of the gas in \dlas\ has been elusive, primarily because the commonly observed resonance-line transitions are relatively insensitive to the physical conditions of the gas. ÊIn principle, the line widths of these lines set an upper limit to the thermal gas temperature but such analysis is challenged by the combination of limited spectral resolution with the multi-component, blended, and saturated nature of the absorption lines.

Direct measurements of cold gas in the Milky Way have previously been carried out using ultra-high resolution studies by, for example, ~\cite{pettini88} and ~\cite{price01}.  ~\cite{pettini88} took advantage of the bright supernova SN1987A and an ultra-high resolution spectrograph on the Anglo-Australian Telescope (AAT).  They observed sodium (Na\ I) gas in the Milky Way Galaxy towards the SN 1987A with a resolution of FWHM = 0.48 \kms , and put limits on the temperature of the gas of T $\leq$ 170 K.  ~\cite{price01} also used the AAT to study interstellar absorption lines towards bright stars in Orion with an instrumental resolution of 0.34 \kms\ FWHM.  In some cases they were able to use multiple species, i.e. Na\ I and calcium (Ca\ II ), to constrain not only the kinetic temperature, but also the turbulent contribution to the broadening of the absorption profile of the gas.  Their excellent resolution and high signal to noise allowed them to measure gas kinetic temperatures in some cases as cold as T = 77 K.  

However, the relatively faint magnitudes of high redshift quasars used to probe \dla\ gas make ultra-high resolution spectroscopy prohibitive.   Several techniques  have been employed to determine the temperature of the \dla\ gas indirectly.  For example, 
 the spin temperature determined from 21 cm absorption towards radio loud quasars is sensitive to the presence of cold gas. 
However, the dearth of detected 21 cm absorbers, which may imply that the gas is warm has lead ~\cite{kanekar03} and ~\cite{kanekar07} to argue that the majority of the gas in \dlas\ is in a Warm Neutral Medium (WNM) where T \rlap{$>$}{\lower1.0ex\hbox{$\sim$}} 1000 K.  However, the spin temperature is uncertain, since it is derived
by assuming that the same {\nhi}  gives rise to both
21 cm and {\lya} absorption. But this is questionable given the
widely different spatial dimensions subtended by the optical
and radio sources at the DLA redshift.  However, ~\cite{kanekar09} argue that the covering factor cannot be the explanation for the generally high spin temperatures.
On the contrary, studies of the UV spectra of \dlas\ have implied the presence, if not predominance, of cold gas: the detection of neutral carbon (\ci ), molecular hydrogen (\htwo ), and in one case CO \citep{srianand08}, which generally require cold temperatures, as well as the upper limit placed on the detection of Si\ II* ~\citep{howk05} have placed tight constrains on the temperatures of the \dla\ gas in several individual cases, while the \ciistr\ technique ~\citep{wolfe03a} has been used to argue that the majority of the gas in \dlas\ \emph{cannot} exist in a WNM without violating the observed bolometric background limit.  

In this paper we present the first \emph{direct} measurement of cold gas in a high redshift \dla .  While our resolution, at FWHM = 6.25 \kms , is not nearly as good as the previously mentioned studies of the Milky Way, we have taken advantage of the many UV multiplets of neutral carbon (\ci\ ) that span both the linear and logarithmic parts of the curve of growth to demonstrate the existence of a sub-1 \kms\ velocity component.  This relatively small Doppler parameter requires a strict upper limit to the thermal temperature of the gas.   

In \S{2} we describe the data, followed by the curve of growth analysis in \S{3}.  In \S{4} we discuss the associated metal lines and molecular hydrogen absorption, followed our conclusions in \S{5}.

\section{The Data}

The well-studied \dla\ 0812$ +$32B at z$_{abs}$ = 2.626 
\citep[see][]{pro03} 
is situated in the line of sight toward a background quasar at z$_{em}$ = 2.701.  This configuration allows for the maximum coverage of \ci\ multiplets not contaminated by blending with the Lyman $\alpha$ forest.  The data were acquired on March 16-17, 2005 with the HIRES echelle spectrograph ~\citep{vogt94} on the Keck I telescope.  A total exposure time of 14,400 seconds using the C1 decker resulted in a signal to noise ratio of $\sim$40 per 1.3 \kms\ pixel with an instrumental resolution of FWHM = 6.25 \kms .  An additional 9,600 second exposure was acquired on January 12, 2008, with the same FWHM, but bluer wavelength coverage and a lower signal to noise ratio (S/N $\sim$ 20, S/N $\sim$ 10 in the blue).  The data were reduced using standard XIDL \footnote{http://www.ucolick.org/$\sim$xavier/IDL/ } reduction packages.  

\section{Analysis}

There are three distinct \ci\ velocity components associated with the \dla\ 0812$ +$32B at velocities of 0 \kms , $-$14 \kms , and $-$57 \kms,   
with respect to an arbitrarily determined z$_{abs}$ = 2.626491.
In this work we focus on the component at v = 0 \kms .  
We measured the equivalent widths (W$_{\lambda}$) and performed a curve of growth analysis on four unblended \ci\ multiplets containing this component.  Because the equivalent width is independent of the instrumental resolution, the curve of growth analysis can reveal relatively narrow velocity components despite the fact that the measured velocity dispersion is well below the instrumental resolution.

\subsection{Curve of Growth Analysis}

We performed the curve of growth analysis on the unblended transitions of the component at v = 0 \kms .  The measured rest-frame equivalent widths over the velocity interval v = [$-$6, 6] \kms\ are given in Table~\ref{tab:EWmorton}.  The equivalent widths were measured by summation of the normalized flux over the given velocity interval on a pixel by pixel basis, i.e.,

\begin{equation}
W_{\lambda} = \sum_{i=1}^{n} [(1 - F_i) \Delta \lambda _i] \cmma
\end{equation}

\noindent  where F$_i$ is the normalized flux at pixel i and $\Delta \lambda _i$ is the pixel width in Angstroms.  The error in the equivalent width was taken to be the standard error as determined by the normalized error array, $\sigma _{F_i}$, as follows:

\begin{equation}
\sigma _{W_{\lambda}} = \sqrt {\sum_{i=1}^{n} (\sigma _{F_i} \Delta \lambda_i)^2 }
\end{equation}

\begin{figure}
\plotone{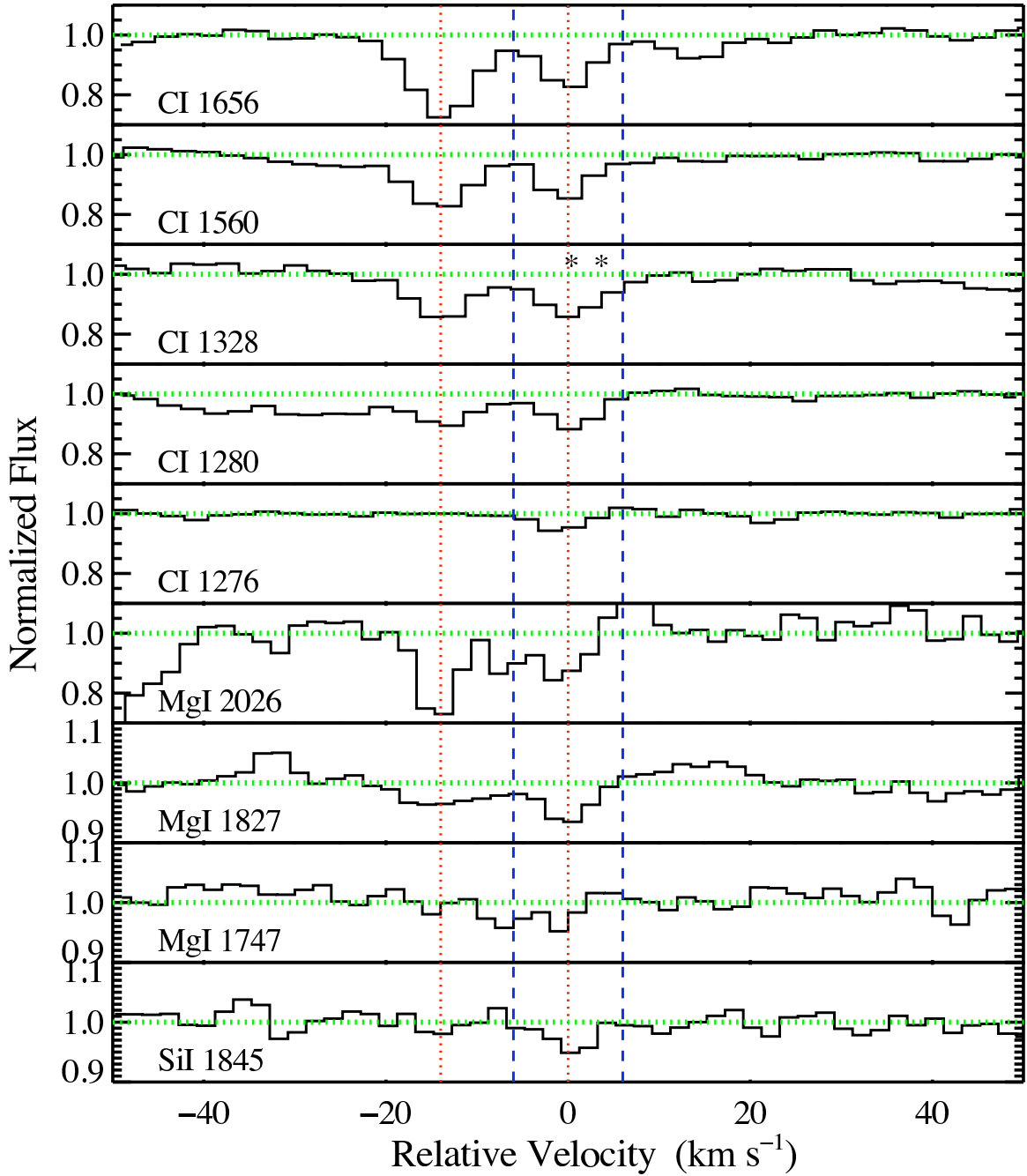}
\caption{ The narrow component, discussed here, is at v = 0 \kms , while a second, stronger component lies at $-$14 \kms\ (both are denoted by [red] dotted lines).  The equivalent width was determined over the velocity interval v = [-6, 6] \kms , denoted in the above figure by the [blue] vertical dashed lines.  The \ci\ $\lambda$ 1328 transition contains blends with \cistr\ transitions from another component, the positions of which are marked here by asterisks.  
}
\label{fig:velplt}
\end{figure}

\noindent  The data includes coverage of all multiplets redward of \ci\ $\lambda$1260 ($\lambda$1260 itself is blended with a strong Si\ II $\lambda$1260 line and therefore gives no information).  We did not include the CI $\lambda$1277.245 transition because it is blended with a relatively strong \cistr\ $\lambda$1277.750 transition. The \ci\ $\lambda$1328 line is weakly blended with two \cistr\ transitions from another component and therefore is an upper limit.  In Figure~\ref{fig:velplt} we plot five \ci\ transitions:  the four \ci\ resonance transitions used to determine the Doppler parameter, b,  where b = $\sqrt{2}$ $\sigma$, as well as the \ci\ $\lambda$ 1328 transition that was excluded because of blending, along with three resonance transitions of Mg\ I and one of Si\ I.  The narrow component is clearly traced by the resonance lines of these other species (Figure~\ref{fig:velplt}), lending confidence to our results.

\begin{figure}
\plotone{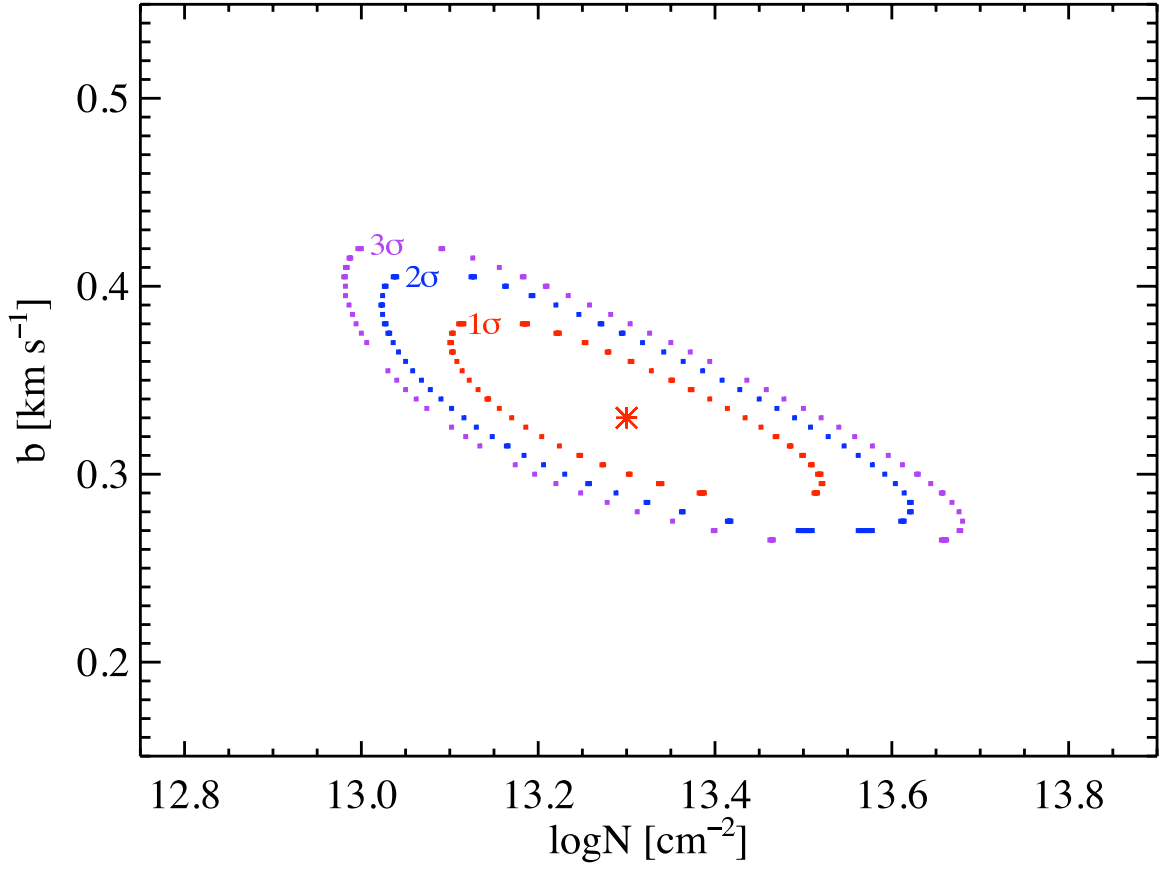}
\caption{ 1$\sigma$, 2$\sigma$, and 3$\sigma$ contours of the $\Delta$$\chi$$^2$ test, using the Morton 2003 atomic data.  The best fit value is logN = 13.30 cm$^{-2}$ and b = 0.33 \kms , which corresponds to a thermal temperature of T $\leq$ 78 K (T $\leq$ 115 K, 2 $\sigma$).   
}
\label{fig:deltachi_morton}
\end{figure}

We performed a $\Delta$$\chi$$^2$ curve of growth test to determine the best fit Doppler parameter, b = 0.33 \kms , and column density, logN(\ci\ ) = 13.30 cm$^{-2}$ \footnote{There is some confusion over the correct f-values for the \ci\ transitions.  If we instead use the \cite{jenkins06} set of f-values we obtain a best fit Doppler parameter b = 0.39 \kms and logN(\ci ) = 12.96 cm$^{-2}$.  Since we cannot determine which set of f-values is more correct, in this paper we use the Morton 2003 values to be consistent with what has previously been used in the literature.}.  
We excluded all upper limits and blends and so performed the test using only the $\lambda$1656, $\lambda$1560, $\lambda$1280, and $\lambda$1276 transitions.  In Figure~\ref{fig:deltachi_morton} we plot the 1, 2, and 3$\sigma$ contours in the column density (N) and Doppler parameter (b) space.  It is apparent that the Doppler parameter b is constrained to be below 1 \kms\ with high significance.  

\begin{figure}
\plotone{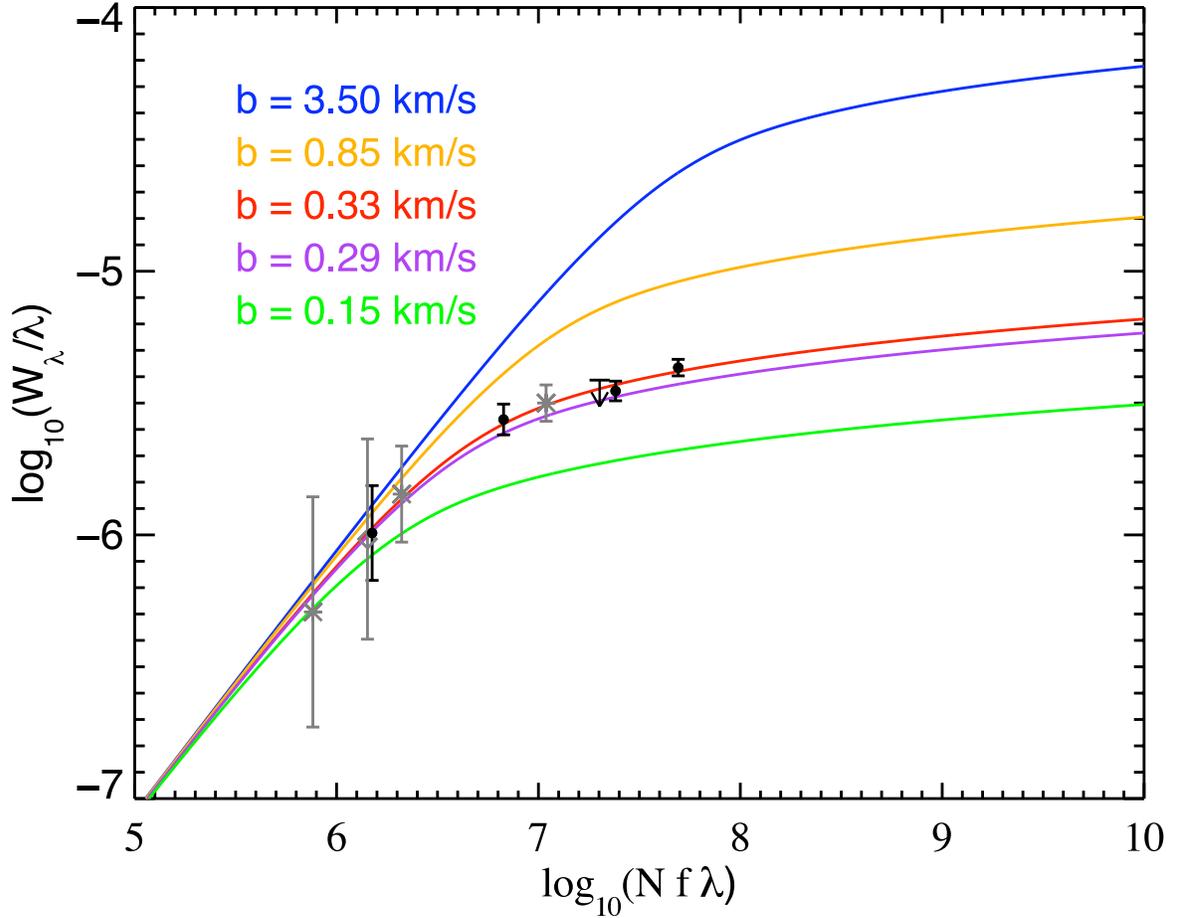}
\caption{ The reduced equivalent width, $\frac{W_{\lambda}}{\lambda}$, versus log (Nf$\lambda$) for the resonance transitions: black circles, from left to right, are \ci\ $\lambda$1276, $\lambda$1280, $\lambda$1328 (upper limit), $\lambda$1560, $\lambda$1656, where log N(\ci ) = 13.30 cm$^{-2}$.  The non-carbon transitions are depicted in gray, where Mg\ I transitions are marked by asterisks and the Si\ I transition by a diamond (from left to right they are: Mg\ I $\lambda$1747, Si\ I $\lambda$1845, Mg\ I $\lambda$1827, and Mg\ I $\lambda$2026, where log N(Si\ I) = 11.53 cm$^{-2}$, and log N(Mg\ I) = 12.68 cm$^{-2}$ as explained in the text).  Five example curves of growth are shown with a range of Doppler parameters.  
}
\label{fig:cog}
\end{figure}

  In Figure~\ref{fig:cog} we plot the reduced equivalent width, $\frac{W_{\lambda}}{\lambda}$, along with representative curves of growth for various Doppler parameters, for illustrative purpose.  The \ci\ data and 1$\sigma$ error bars are overplotted as black points.  It is apparent that neither the b = 3.5 \kms , nor the b = 0.85 \kms\ curves of growth, are good fits to the data.  The small Doppler parameter is independently confirmed by the analysis of the Mg\ I transitions, for which a $\Delta$$\chi$$^2$ test determines a best fit of b = 0.35 \kms , and log N(Mg\ I) = 12.68 cm$^{-2}$.  These points are overplotted as gray asterisks.  The gray diamond represents the Si\ I line, measured using the VPFIT package, version 9.5 \footnote{http://www.ast.cam.ac.uk/$\sim$rfc/vpfit.html}, because it is a single transition and therefore we could not perform an independent curve of growth test.  Fixing the Doppler parameter to that of \ci , scaled by the square root of the ratio of their atomic masses, results in b = 0.22 \kms\ and log N(\ion{Si}{1}) =  11.53 $\pm$ 0.18 cm$^{-2}$.  
The positive detection of \ion{Si}{1} is extremely rare in DLAs and therefore
strongly supports the presence of a dense, cold medium.

It is important to note that we are able to constrain the Doppler parameter of the \ci\ component because the data are approaching the saturated portion of the curve of growth.  This can be seen both from the values of $\tau$, the optical depth, given in Table~\ref{tab:EWmorton}, as well as from the model curves of growth shown in Figure~\ref{fig:cog}.  This is perhaps best illustrated by calculating the ratio of the reduced equivalent widths of two transitions, for example, $\lambda$1656 and $\lambda$1280.  
On the linear part of the curve of growth, we expect the ratio of the reduced equivalent widths to increase linearly according to the ratios of f$\lambda$, i.e.
\begin{equation}
\frac{(\frac{W_{\lambda}}{\lambda})^{1656}}{ (\frac{W_{\lambda}}{\lambda})^{1280}} = \frac {  (f\lambda)^{1656} }{  (f\lambda)^{1280}} \approx 4
\end{equation}
\noindent Even if we allow a 2$\sigma$ shift in opposite directions from the measured values, we still only obtain a ratio of equivalent widths equal to 2.35, rather than the expected 4.  It is clear that both points are not on the linear part of the curve of growth and one can therefore determine a unique b value.  

\begin{figure}
\plotone{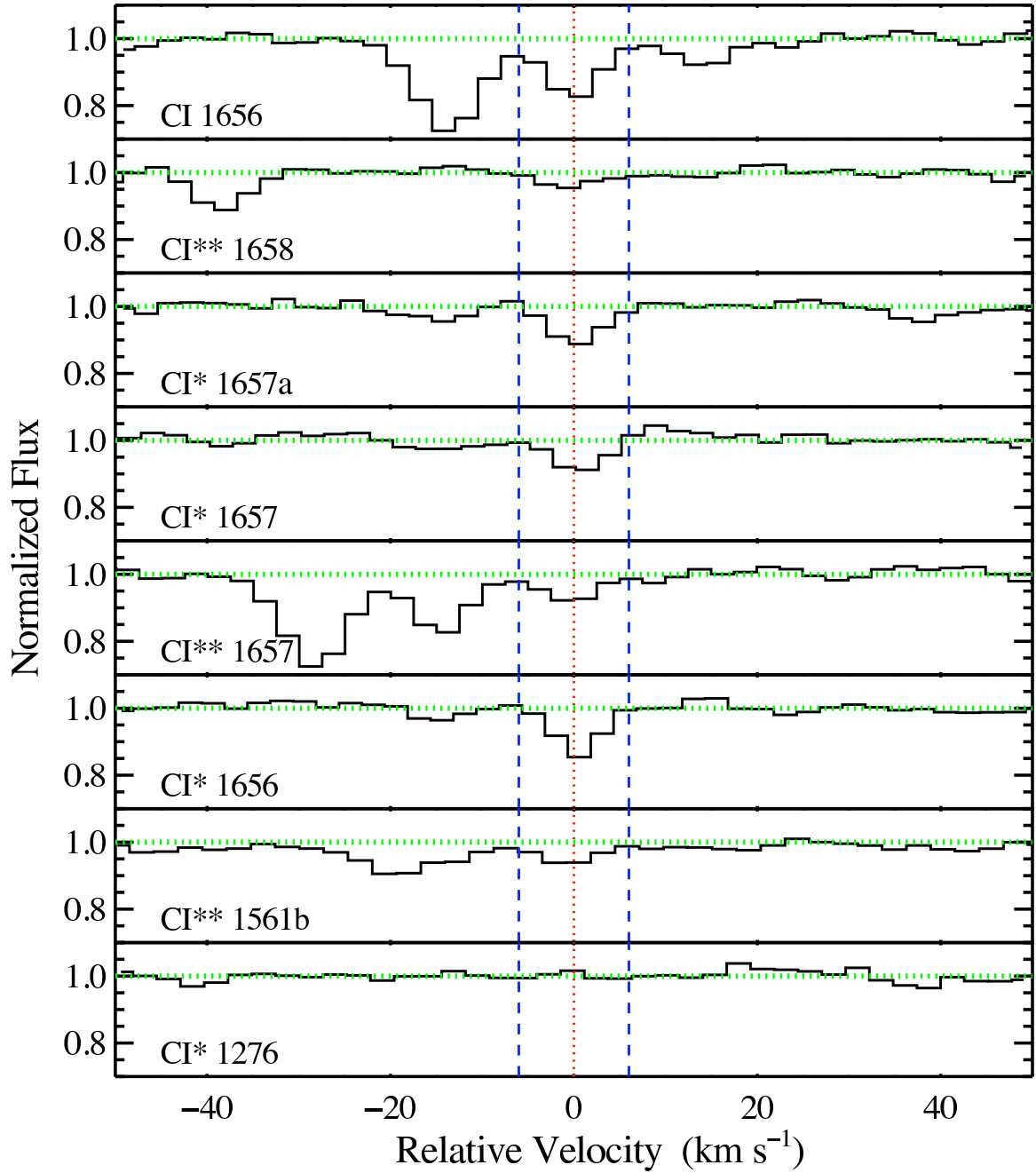}
\caption{ The unblended \ci\ fine structure transitions.  The narrow component, discussed here, is at v = 0 \kms .  The equivalent width was determined over the velocity interval v = [-6, 6] \kms , denoted in the above figure by the [blue] vertical dashed lines.  
}
\label{fig:velplt_fnstrct}
\end{figure}

\begin{figure}
\plotone{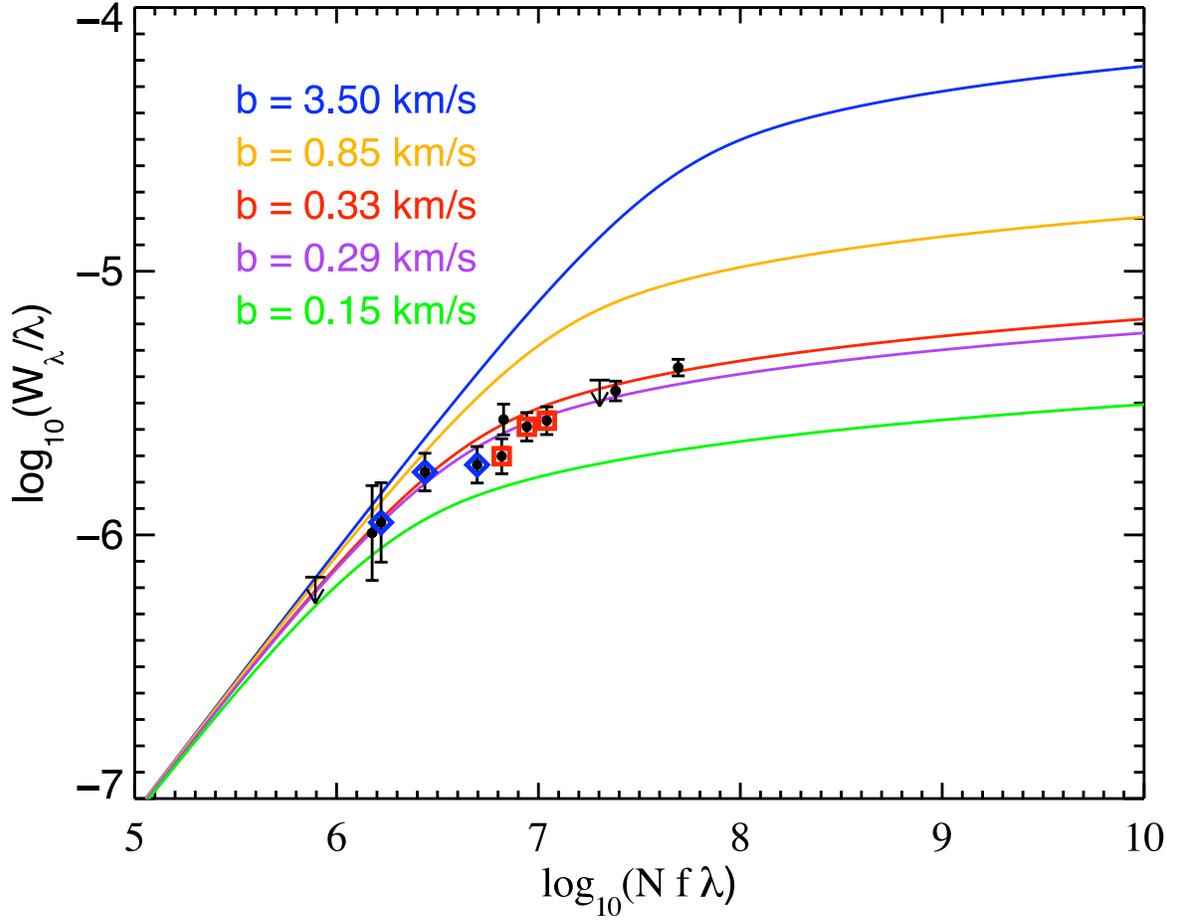}
\caption{ The reduced equivalent width, $\frac{W_{\lambda}}{\lambda}$, versus log (Nf$\lambda$) for the resonance transitions: black circles, from left to right, are \ci\ $\lambda$1276, $\lambda$1280, $\lambda$1328 (upper limit), $\lambda$1560, $\lambda$1656.  The unblended \ci\ fine structure transitions are denoted by diamonds (\cistrstr ) and squares (\cistr ).  They are, from left to right, \cistr\ $\lambda$1276 (marked as upper limit), \cistrstr\ $\lambda$1658, \cistrstr\ $\lambda$1561, \cistrstr\ $\lambda$1657, \cistr\ $\lambda$1657.3, \cistr\ $\lambda$1657.9, and \cistr\ $\lambda$1656.  Five example curves of growth are shown with a range of Doppler parameters.  
}
\label{fig:cog_fnstrct}
\end{figure}

Further evidence in support of the narrow component is its presence in several of the associated \ci\ fine structure transitions, denoted \cistr\ and \cistrstr .  While many of these are blended with either close transitions or transitions from other components, several are relatively unblended.  We show these unblended transitions in Figure~\ref{fig:velplt_fnstrct}.  In Figure~\ref{fig:cog_fnstrct} we plot these transitions on the curve of growth, using the equivalent width as measured over the same velocity width, v = [-6, 6] \kms , along with the \ci\ resonance transitions.  Table~\ref{tab:EWmorton} lists the measured equivalent width for each fine structure transition used, and Table~\ref{tab:fit} gives the column densities resulting from fitting of these lines using VPFIT assuming a fixed Doppler parameter of b = 0.33 \kms .  
The excitation temperature of the gas, implied by the ratios of the fine structure level populations and the Boltzmann equation are as follows, $ \frac{\mcistr\ }{\ci\ } $ = 12.70 K, $ \frac{\mcistrstr\ }{\ci\ } $ = 16.65 K, and $ \frac{\mcistrstr\ }{\mcistr\ } $ = 20.56 K.  These are well above the expected temperature of the Cosmic Microwave Background Radiation (CMB) at the \dla\ redshift (T$_{CMB}$ = 2.725 (1 + z$_{abs}$) = 9.88 K), implying that, as expected, the CMB is not the only excitation mechanism of the \ci -bearing gas.  

The significance of an unambiguously narrow Doppler parameter is that it requires a strict upper limit on the temperature of the gas.  For a given absorption line, the Doppler parameter b is defined as b$^2$$_{total}$ = b$^2$$_{thermal}$ + b$^2$$_{turbulence}$.  If we ignore the turbulent contribution, i.e. set b$_{turbulence}$ = 0, we can write b$^2$$_{total}$ = b$^2$$_{thermal}$ = $\frac{2kT}{m}$, where k is Boltzmann's constant and m is the mass of the particle.  This can be rewritten as b$^2$$_{thermal}$ = $\frac{T}{60A}$, where A is the atomic weight of the element.  The assumption that the turbulent contribution is negligible provides the upper limit on the thermal temperature, as any turbulent contribution would lead to a decrease in the thermal contribution.  Using the best fit Doppler parameter of b = 0.33 \kms , and A = 12 for carbon, results in T$_{thermal}$ $\leq$ 78 K.  The 2$\sigma$ upper limit of b = 0.40 \kms provides a 2$\sigma$ upper limit of T$_{thermal}$ $\leq$ 115 K.    
 
 To our knowledge this is the first direct measurement of cold gas in a high redshift \dla\ galaxy (however, also see Carswell et al., 2009, in preparation, on a narrow, cold component in \dla\ 1331$+$17).  The measurement of the small, sub-resolution Doppler parameter was made possible by the application of the standard curve of growth technique using several multiplets of the generally weak neutral carbon absorption line.  An obvious next step would be to obtain ultra high resolution spectra (FWHM $\sim$ 0.25 \kms ) in order to resolve the narrow component and fully map the velocity structure of the gas.  This goal is perhaps unattainable with current telescopes given the relatively faint magnitude of the high redshift quasar (R $\sim$ 17.7), and therefore prohibitive exposure time, and must await the next generation of large telescopes and high resolution spectrographs.

\subsection{Systematic Effects and Assumptions}

Our curve of growth analysis relies on two fundamental assumptions.  
We now detail each and the impact on our results.

\subsubsection{Single Component}

The first assumption is that the absorption can 
be well modeled by a single component at v = 0 \kms .  
If there are actually two or more components, for example a strong, narrow component near to a broader, weak component, we could underestimate the total column density in the narrow component (see for example \cite{nachman73, pro06}).  However, this scenario would require that we simultaneously overestimate the Doppler parameter of the narrow component and thus, our upper limit on the thermal temperature encompasses any such possibility.  

\subsubsection{Partial Covering}

The second assumption is that the narrow component absorption is not an artifact of partial covering of the background quasar.  If the \ci\ cloud is physically small, on the order of a few parsecs, it could be partially covering the background quasar.  This scenario could mimic the observed narrow component because the partial covering would cause a decrease in the observed equivalent width of the absorption lines.  If the true equivalent width of each transition were actually larger, all of the points on the curve of growth would be shifted upwards, while maintaining their relative positions.  However, because the column density, N, is a free parameter, N could also increase, shifting the points to the right, thus allowing a larger Doppler parameter. We note, however, that if the \ci\ is physically associated with the molecular hydrogen (\htwo ) in the same velocity component -- which is the common assumption -- partial coverage is \emph{not} a possibility in this case because, as we show in the next section, the saturated \htwo\ J = 0 and J = 1 lines have zero central intensities and therefore rule out
the partial covering scenario.

\subsection{Molecular Hydrogen}

Further evidence in support of the presence of cold gas is the detection of strong molecular hydrogen (\htwo ) absorption in DLA 0812$+$32B.  \htwo\ gas is expected because \ci\ and \htwo\ are photoionized and photodissociated respectively by photons of the same energies and therefore, while rare in absorption line systems overall, 
when observed, they are usually found together \citep{noter07}.  
In this \dla , the velocity profile of the \htwo\ gas is consistent with that of the \ci\ -- three velocity components at $v = (0, -14, -57)$\kms\ -- and in the following we will discuss only the \htwo\ component associated with the narrow \ci\ component at z$_{abs}$ = 2.626491.

We used VPFIT to measure the column densities of the rotational J states of \htwo\ associated with the narrow \ci\ component.  The \htwo\ J = 0 and J = 1 states are damped and thus provide little constraint on the Doppler parameter.  Assuming that the gas is associated with the narrow \ci\ component, we fixed the Doppler parameter to that of \ci , scaled by the square root of the ratios of their masses, i.e. b$_{H_2  }$ =  (6)$^{1/2}$b$_{C I  }$ = 0.81 \kms .  We detected and measured the \htwo\ J rotational states up to and including the J=4 state and we give the resultant column densities in Table~\ref{tab:htwo}. 
  
 We emphasize two key results from the analysis of the \htwo\ that support our conclusions of cold gas in the narrow \ci\ component.  
 First, we estimate the kinetic temperature of the gas using the column densities of \htwo\ in the J=0 and J=1 rotational states and assuming the states are thermalized according to the Boltzmann distribution (see equation 8 in ~\cite{levshakov02}).  The excitation temperature, T$_{ex}$, is defined by,  

\begin{equation}
\frac{N(J)}{N(0)} = \frac{g(J)}{g(0)} e^-{\frac{B_v J (J + 1)}{T_{ex}}}
\end{equation} 

\noindent where B$_v$ = 85.36 K for the vibrational ground state and g(J) is the degeneracy of level J.  The typical assumption is that the kinetic temperature of the gas can be estimated by the excitation temperature derived from the population of \htwo\ in the states J = 0 and J = 1, assuming that the J = 1 level is thermalized.  For the molecular gas associated with the narrow component, we derive T$_{ex}$ = 46 K.  This temperature is consistent with that derived from the \ci\ data, thus providing additional evidence that not only are the \htwo\ and the \ci\ physically associated, but also, the gas temperature is remarkably cold.  We stress that because the J=0 and J=1 states are damped, their measured column densities are not affected by changes in the Doppler parameter, and therefore this analysis is independent of the choice of \htwo\ Doppler parameter over the range consistent with the \ci\ value (i.e. b \rlap{$<$}{\lower1.0ex\hbox{$\sim$}} 1 \kms ).

\begin{figure}
\plotone{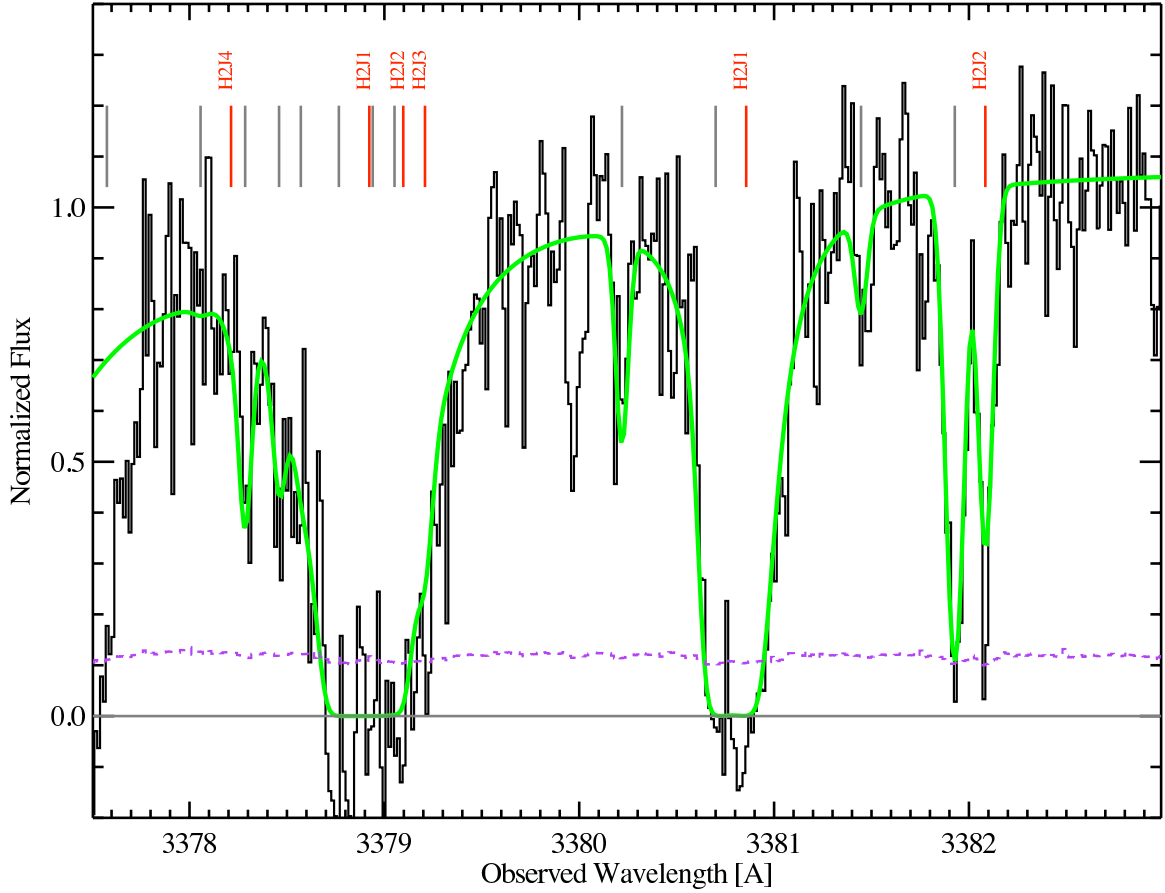}
\caption{ A section of the spectrum taken January 2008, showing several transitions of molecular hydrogen (\htwo\ ).  The flux is shown in black while the VPFIT derived fit is overplotted in [green] gray and the error array is a dashed [purple] line.  The \htwo\ transitions corresponding to the narrow component are labeled [in red].  The \htwo\ transitions corresponding to the other \ci\ components are marked by gray dashed lines.  This section of spectrum was chosen because it is free from blending with lyman $\alpha$ forest lines and therefore easily seen that the \htwo\ transitions are black at line center, thus precluding the possibility of partial covering of the background quasar.  }
\label{fig:h2}
\end{figure}

The second result that we derive from an analysis of the \htwo\ data is that the damped J = 0 and J =1 transitions are black at line center.  This would not be the case if there was partial covering of the background quasar, as light leakage would produce a non-zero baseline.  Because we do not see evidence of this in the many saturated \htwo\ transitions -- see for example the section of spectrum shown in Figure~\ref{fig:h2} -- we argue that partial covering of the background quasar cannot explain the small Doppler parameter of the \ci\ gas.  

Finally, it is interesting to note that the amount of \htwo\ in this component, log N(\htwo )$_{total}$ = 19.88 cm$^{-2}$, makes this one of the most molecule-rich, high redshift \dlas .  The fraction of \htwo , defined as f$_{H_2} = 2N(H_2)/[N(H I) + 2N(H_2)] $ $\geq$ 0.06, is a lower limit because of the uncertainty as to how much of the N(H I) is associated with the narrow component.  This is yet another piece of evidence of the presence of cold gas.

\section{Discussion}

\subsection{Associated Metals}
\begin{figure}
\plotone{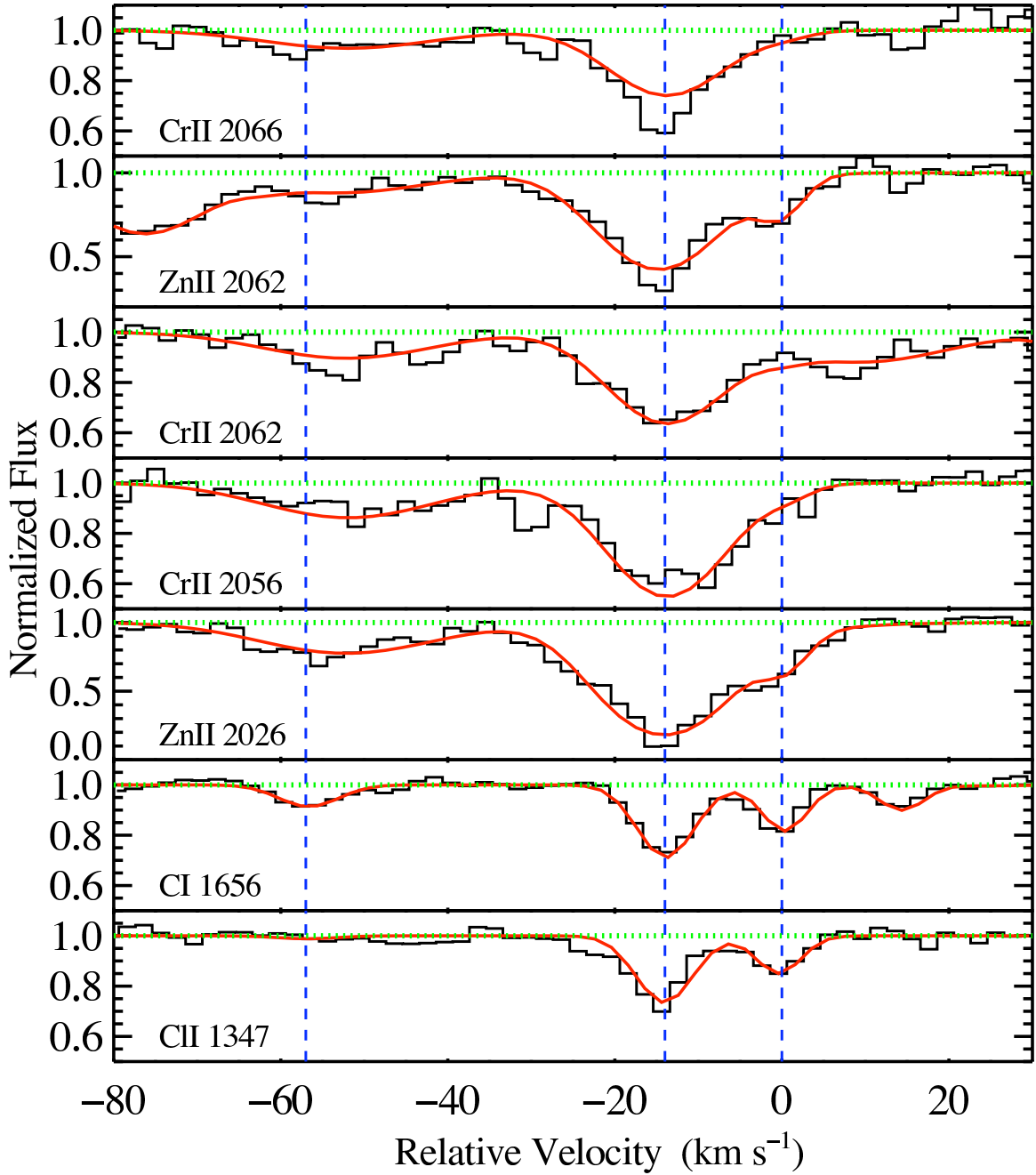}
\caption{ Low ions indicating a high level of dust depletion in the narrow velocity component, depicted here at v = 0 \kms . Note that the component is well traced by Cl I $\lambda$1347 and Zn\ II but not by Cr II, indicating high levels of dust depletion in the narrow velocity component.  Vertical dashed lines indicate the positions of the three velocity components.    
}
\label{fig:low_ions}
\end{figure}

In Figure~\ref{fig:low_ions} we compare the detailed velocity structure of the \ci\ with that of the other low-ions, in order to explore the metal distribution of the narrow component.  It is apparent that while the \ci\ velocity structure is generally well traced by the low-ions such as Zn\ II and Cr\ II and the resonance transition chlorine (Cl\ I ), there is an obvious lack of Cr II in the narrow velocity component, indicative of strong depletion by dust.  Because of blending with the component at $-$14 \kms , we were not able to measure the equivalent widths of the associated metal lines.  Instead, we used VPFIT to simultaneously fit the three velocity components and measure the associated metal column densities.  The fit is indicated by the red line in Figure~\ref{fig:low_ions} and the resulting column densities are given in Table~\ref{tab:low_ions}.  

If real, the resultant measurement of log N(Zn\ II) = 13.7 cm$^{-2}$ implies that this is one of the most metal-rich high redshift systems and the Zn to Cr ratio implies a dust to gas ratio of $\sim$ 40\% of the Milky Way, compared with typical \dla\ dust to gas ratios of 1/30 solar.  However, this extreme metallicity is contradicted by the measurement limits on other elements like oxygen.  Assuming solar relative abundances and given log N(Zn\ II) = 13.7 cm$^{-2}$, we would expect to measure log N(O\ I) $\approx$ 17.8 cm$^{-2}$.  This is inconsistent with the data as we can place a 3 $\sigma$ upper limit of log N(O\ I) $<$ 17.07 cm$^{-2}$, corresponding to a measured equivalent width of O\ I $\lambda$1355, W$_{\lambda }$ = 0.50 $\pm$ 0.74 mA.  While the Zn\ II data seem to indicate that the metal content of the narrow \ci\ component is high, this model implies an extreme environment that is not consistent with other elements assuming solar relative abundances.  In addition, careful inspection of the data reveal that there appears to be a slight offset in velocity between the Zn\ II and \ci\ components, on the order of 1 \kms .  This data alternatively can be explained by the presence of an additional weak, broad Zn\ II component at approximately the same redshift.  If the Zn\ II narrow component contains the maximum amount of Zn\ II as allowed from the upper limit on O\ I, log N(Zn\ II) = 13.0 cm$^{-2}$, the additional broad component is required to contain log N(Zn\ II) = 12.32 $\pm$ 0.24 cm$^{-2}$ at z = 2.626447  $\pm$  0.000031 with b = 5.6 $\pm$ 3.3 \kms . 
We conclude that while the exact distribution of the low ion metals in the narrow component is uncertain, it is clear that the narrow component likely contains a higher dust content, an expected result given that the narrow component is colder and most likely denser than the surrounding gas.  
   
  \subsection{Gas Density, Pressure, Gravitational Confinement}
  
  While it may be unclear exactly how the metals are distributed with respect to the cold \ci\ component, it is clear that this cold component has associated \ci\ fine structure lines, which provide additional information about the gas.  The volume density and temperature of the gas can be constrained by analysis of the \ci\ fine structure lines, given in Table~\ref{tab:EWmorton} (for example, see Jorgenson et al., 2009, in prep and ~\cite{noter07}).  We solved the steady state equation for each fine structure level, assuming T = 100 K, and including collisions with neutral hydrogen, electrons and protons, as well as excitation by the CMB, the Haardt-Madau background and the local radiation field, as measured by the \ciistr\ technique ~\citep{wolfe03a}, to estimate the volume density of this cloud.  This results in a range of densities 
n(\hi ) $\sim 40 - 200$cm$^{-3}$ (2$\sigma$) from which we can estimate the pressure of this cloud, P/k = nT.  Therefore, we estimate P/k $\sim 2500 - 16,000$cm$^{-3}$ K, slightly higher than the pressures derived from \ci\ gas in the Milky Way where the median P/k = $\sim$2240 cm$^{-3}$ K ~\citep{jenkins01}.   
For comparison, giant molecular clouds in the disc of the Milky Way, M64 and the Antennae galaxy span a range of P/k = $\sim$10$^5$ $-$ 10$^7$cm$^{-3}$ K \citep{tate2004, ros2007}.  
  
  We can use the measured volume density of the narrow component to estimate the physical size of the gas cloud.  However, because the HI Lyman-$\alpha$ line is damped, thus hiding any information about its velocity structure, we do not know how much of the neutral hydrogen column density is associated with the narrow component cloud.  Instead, we make the assumption that the the column density of \htwo , whose weaker non-damped lines clearly indicate a velocity structure matching that of \ci , can be used as a lower limit to N(\hi ) associated with the cloud.
  Using the volume density estimated from the \ci\ fine structure data gives a length, 
$\ell\ $ $\geq$ N(\htwo\ )/n(\hi ) = 10$^{19.88}$ cm$^{-2}$/(40 - 200 cm$^{-3}$)  
= (1.90 - 0.38) $\times$10$^{18}$ cm $\approx$ (0.6 - 0.1) pc.  
Therefore, we set a lower limit to the size of the cloud of $\sim 0.1$ parsec. 
    
  Given this physical picture, we are tempted to consider if such a cloud is gravitationally confined.  We estimate the gravitational weight per unit area, W,  of the cloud by, 
  
  \begin{equation}
  W = \frac{\pi\ G \Sigma\ ^2}{2} \cmma
  \end{equation}    
  
  \noindent where G is the gravitational constant and $\Sigma$ is the gas surface density given by, 
  
  \begin{equation}
  \Sigma\ = \mu\ m_H  N_{HI}^{perp}  \perd 
  \end{equation}
    
\begin{figure}
\plotone{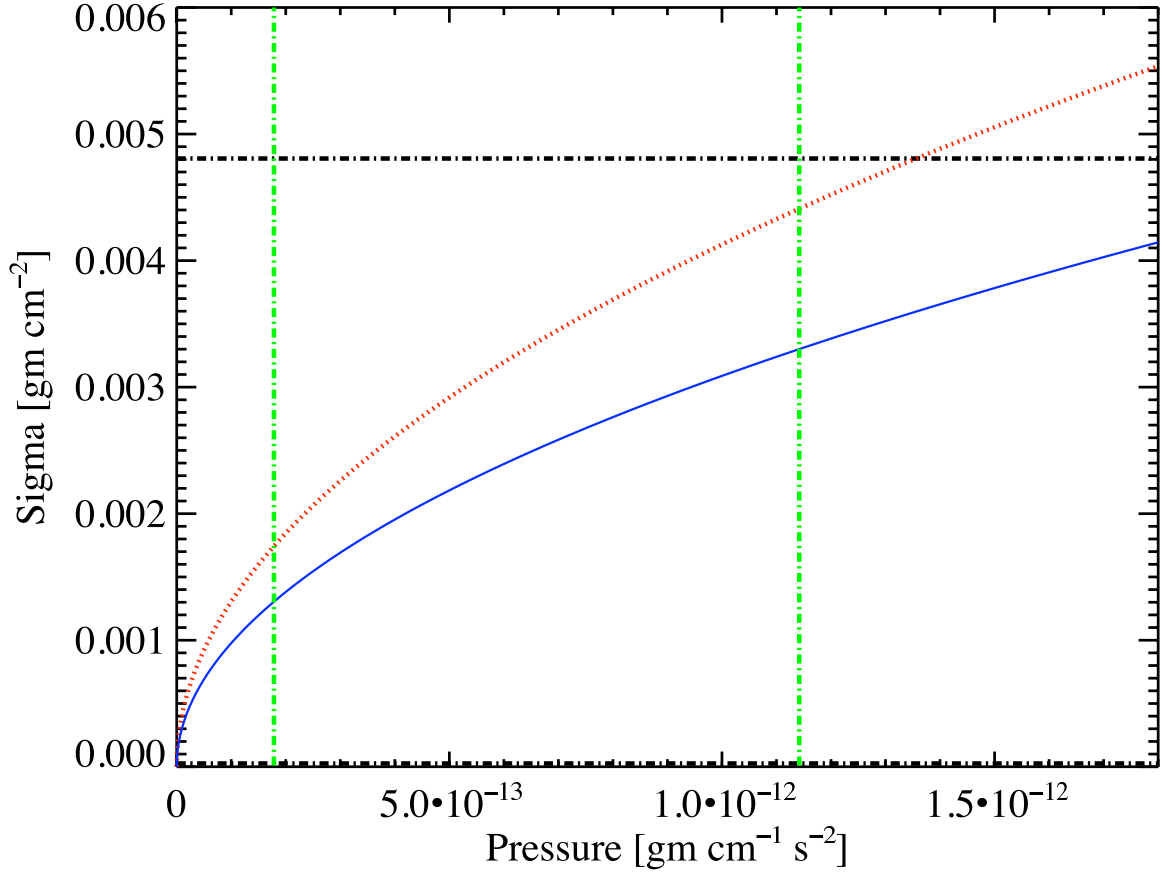}
\caption{ $\Sigma$ versus pressure for a range of volume densities.  The horizontal dot-dashed lines indicate the allowed range of $\Sigma$ which includes 95 \% of the inclination angles and is centered on the average, cos(i) = 60, for two cases -- the first assumes the full measured log N$_{HI}^{observed}$ = 21.35 cm$^{-2}$ and creates the upper bound at $\Sigma$ $\sim$ 4.8$\times$10$^{-3}$ gm cm$^{-2}$.  The second assumes a lower limit of log N$_{HI}^{observed}$ = 19.88 cm$^{-2}$ (determined by molecular hydrogen) and gives the lower bound at $\Sigma$ $\sim$ 3$\times$10$^{-5}$ gm cm$^{-2}$.  The [red] dotted line indicates the pressure and weight equality for a spheroidal model, while the [blue] solid line indicates the same for the plane parallel geometry.  Clouds falling in the region above the lines are self-gravitating.  The [green] vertical dashed lines indicate the 2$\sigma$ density range allowed by the \ci\ fine structure data.  It is apparent that there is up to 50\% probability that the cloud is gravitationally rather than pressure confined. }
\label{fig:sigma}
\end{figure}

  \noindent We estimate the N$_{HI}^{perp}$ for a range of disk inclination angles, i,  assuming N$_{HI}^{perp}$ = N$_{HI}^{observed}$ cos(i).  
In Figure~\ref{fig:sigma} we plot $\Sigma$ versus the thermal pressure, P, that we estimate as P = 1.3 m$_H$n(\hi )$\sigma$$^2$, where n(\hi ) is the volume density and $\sigma$ is the velocity dispersion, $\sigma _{\hi }$ = 0.81 \kms\ \footnote{Derived from b$_{\ci }$ = 0.33 \kms = $\frac{b_{\hi }}{\sqrt{12}}$  and $\sigma$ = $\frac{b}{\sqrt{2}}$ }.   
The horizontal dot-dashed lines indicate the allowed range of $\Sigma$ which includes 95 \% of the inclination angles and is centered on the average, cos(i) = 60, for two cases -- the first assumes the full measured log N$_{HI}^{observed}$ = 21.35 cm$^{-2}$ and creates the upper bound at $\Sigma$ $\sim$ 4.8$\times$10$^{-3}$ gm cm$^{-2}$.  The second assumes the lower limit of log N$_{HI}^{observed}$ = 19.88 cm$^{-2}$ , as determined by the molecular hydrogen, and gives the lower bound at $\Sigma$ $\sim$ 3$\times$10$^{-5}$ gm cm$^{-2}$.
The dotted red line indicates the pressure and weight equality for a spheroidal model, while the solid blue line indicates the same for the plane parallel geometry.  Gravitational confinement is indicated by the region above these lines, i.e. where the weight of the gas exceeds the turbulent pressure.  It is apparent that, depending upon an unknown quantity -- how much of the N(\hi ) is associated with the cold component -- for the range of densities determined by the \ci\ fine structure data, indicated by the green vertical lines, there is up to 50\% probability that the gas is gravitationally confined.  This is significant given that the gas in typical \dlas\ is not gravitationally bound, but rather pressure confined.  
We propose that this \ci\ cloud (and perhaps other \ci\ clouds) could be tracing an evolutionary stage of \dla\ gas, from primarily neutral atomic gas, to a colder, denser, primarily molecular phase, and hence, possibly the birth of what will eventually become a region of star formation.

  \section{Conclusions}
  
  We have presented the first direct evidence of cold gas in a high redshift galaxy.  We used the curve of growth on several multiplets of the ground state of neutral carbon to derive an unambiguously narrow Doppler parameter, b = 0.33 \kms , corresponding to a 2 $\sigma$ upper limit on the thermal gas temperature of T $\leq$ 115 K.  We have measured the Lyman and Werner-band absorption bands of molecular hydrogen resulting in a log N(\htwo )$_{total}$ = 19.88 cm$^{-2}$ and a fraction of molecular hydrogen, f$_{_2} $ $\geq$ 0.06.  Using the damped J = 0 and J = 1 molecular hydrogen lines we determined an excitation temperature of T$_{ex}$ = 46 K consistent with the results derived from the \ci\ data.  We have used the associated \ci\ fine structure lines to determine the volume density of the cloud to be n(\hi ) $\sim 40 - 200$ cm$^{-3}$ (2$\sigma$), resulting in a lower limit on the cloud size of approximately 0.1 $-$ 1 parsec.  This small size indicates that we are likely probing a small dense clump within the larger \dla , rather than diffuse gas filing the \dla .  We speculate that this cloud may be gravitationally confined and may be a tracer of the transition region from primarily atomic to primarily molecular gas and the sites of star formation at high redshift.  

\acknowledgments
RAJ acknowledges support from the STFC-funded Galaxy Formation and Evolution programme at the Institute of Astronomy, University of Cambridge.  RFC is grateful to the Leverhulme Trust for an Emeritis award.  
The authors wish to extend special thanks to those of Hawaiian ancestry on whose sacred mountain we are privileged to be guests.  Without their generous hospitality, none of the observations presented in this work would have been possible.

\clearpage

  \begin{deluxetable}{lcccc}
\tablewidth{0pc}
\tablecaption{EQUIVALENT-WIDTH MEASUREMENTS\label{tab:EWmorton}}
\tabletypesize{\footnotesize}
\tablehead{
\colhead{transition} & \colhead{$\lambda$ } & \colhead{$f$} & \colhead{W$_{\lambda}$} &\colhead{$\tau _{0}^b$}\\
&[\AA\ ]&&[m\AA\ ]
}
\startdata
\ci\ &1656.9284 &  0.14900  &  7.14    $\pm$ 0.50&27.9\\
\ci\ &1560.3092 &  0.07740 &5.48   $\pm$   0.45&13.6\\
\ci\ &1328.8333 &  0.07580    &5.13 $^c$  $\pm$   0.43&...\\
\ci\ &1280.1352 &  0.02630    &3.51   $\pm$   0.44&3.8\\
\ci\ &1276.4822 &  0.00589    &1.30   $\pm$  0.44&0.85 \\
Mg\ I&2026.4768   & 0.11200& 6.41 $\pm$ 2.31&4.7\\
Mg\ I&1827.9351&0.02390 & 2.61 $\pm$ 0.56&0.9\\
Mg\ I&1747.7937& 0.00908& 0.89 $\pm$ 0.60&0.3\\
Si\ I&1845.5200&0.22900&1.78 $\pm$ 0.67&0.7\\
\hline
\hline
\cistrstr\ &1658.1211 & 0.03710 &   1.85 $\pm$ 0.54&0.9\\
\cistr\ &1657.9071& 0.04940& 4.26  $\pm$ 0.50&4.5\\
\cistr\ & 1657.3792& 0.03710 & 3.29 $\pm$   0.47 &3.4\\
\cistrstr\ &  1657.0081 &0.11100 & 3.06$\pm$ 0.45 &2.6\\
 \cistr\ &1656.2672 & 0.06210 &  4.49 $\pm$  0.51 & 5.7\\
\cistrstr & 1561.4378 & 0.06490 & 2.7 $\pm$ 0.41 &1.4\\
 \cistr\ & 1276.7498 & 0.00308 & $\leq$0.88$^d$ & ...\\
\enddata
\tablenotetext{a}{Atomic data is from Morton 2003}
\tablenotetext{b}{Peak optical depth assuming best fit values b$_{\ci\ }$ = 0.33 \kms\ and logN(\ci\ ) = 13.30 cm$^{-2}$, logN(\cistr\ ) = 13.02 cm$^{-2}$, logN(\cistrstr\ ) = 12.47 cm$^{-2}$; logN(Mg\ I) = 12.68 cm$^{-2}$ for b$_{Mg\ I }$ = 0.35 \kms\, ; logN(Si\ I) = 11.53 cm$^{-2}$ for b$_{Si\ I }$ = 0.22 \kms\ }
\tablenotetext{c}{Upper limit due to blending}
\tablenotetext{d}{2$\sigma$ upper limit}
\end{deluxetable}

 \begin{deluxetable}{lcc}
\tablewidth{0pc}
\tablecaption{\ci\ FINE STRUCTURE COLUMN DENSITIES$^{a}$\label{tab:fit}}
\tabletypesize{\footnotesize}
\tablehead{\colhead{transition}&\colhead{b} &\colhead{log N}\\
 & [\kms ]& [cm$^{-2}$]
 }
 \startdata
 \ci\   & 0.33&13.30$^{b}$ \\
 \cistr\ &  0.33&13.02 $\pm$0.03\\ 
\cistrstr\   & 0.33&12.47 $\pm$ 0.05\\ 
\enddata
\tablenotetext{a}{z$_{abs}$ = 2.626491}
\tablenotetext{b}{Best-fit column density as determined by $\Delta$$\chi$$^2$ test. }
\end{deluxetable}

 \begin{deluxetable}{lcc}
\tablewidth{0pc}
\tablecaption{MOLECULAR HYDROGEN ASSOCIATED WITH NARROW COMPONENT \label{tab:htwo}}
\tabletypesize{\footnotesize}
\tablehead{\colhead{transition}&\colhead{b} &\colhead{log N}\\
 & [\kms ]& [cm$^{-2}$]
 }
 \startdata
 H2 (J=0)& 0.81$^a$& 19.79$\pm$ 0.01\\ 
H2 (J=1)&  0.81&19.15$\pm$ 0.01\\ 
H2 (J=2)& 0.81& 16.60$\pm$ 0.03\\ 
H2 (J=3)& 0.81&  15.11$\pm$  0.05\\
H2 (J=4)& 0.81& 13.99$\pm $ 0.04\\
\enddata
\tablenotetext{a}{Doppler parameter tied to that of \ci .  Note that the J = 0 and J = 1 transitions are heavily saturated and therefore the resultant log N is determined by the damping wings and is insensitive to the choice of b.  }
\end{deluxetable}

 \begin{deluxetable}{cccc}
\tablewidth{0pc}
\tablecaption{DUST-TO-GAS RATIO COMPONENT ANALYSIS OF \dla\ 0812$ +$32B\label{tab:low_ions}}
\tabletypesize{\footnotesize}
\tablehead{
}
\startdata
& comp. v = $-$57\kms\ &comp. v = $-$14\kms\ &comp. v = 0 \kms\ $^a$\\
&z$_{abs}$=2.625863 &z$_{abs}$=2.626320 &z$_{abs}$=2.626491\\
&b = 14.38$\pm$1.32 [\kms ] &b = 8.24$\pm$0.31 [\kms ] &b = 0.33$^b$ [\kms ]\\\hline\hline
N(Cr II) [cm$^{-2}$]&12.835    $\pm$0.042& 13.241    $\pm$0.017&11.852   $\pm$ 0.303\\
N(Zn\ II) [cm$^{-2}$]&12.383    $\pm$0.038&13.020    $\pm$0.014 &13.757    $\pm$0.170\\
$[Cr/Zn]$& $-$0.568$\pm$0.057&$-$0.799$\pm$0.022&$-$2.925$\pm$0.347\\
\enddata
\tablenotetext{a}{This is the narrow component analyzed in this paper. }
\tablenotetext{b}{Doppler parameter fixed to match that of \ci\ determined by the $\Delta$$\chi$$^2$ test. }
\end{deluxetable}

\end{document}

%% file: xavier_defs.tex
\def\aap{A \& A}

\def\apj{ApJ}

\def\apjs{ApJS}

\def\araa{ARAA}
\def\mnras{MNRAS}

\def\lya{Ly$\alpha$ }

\def\etal{et al. }
\def\kms{km~s$^{-1}$ }

\def\s-1{s$^{-1}$}
\def\Hz-1{Hz$^{-1}$}

\def\sci#1{{\; \times \; 10^{#1}}}

\def\l1l2{\lambda_2 \; {\rm and} \; \lambda_1}

\def\intl{\int\limits}

\def\eht1{{\rm \hat e_1}}

\def\ltk{\left [ \,}
\def\ltp{\left ( \,}
\def\ltb{\left \{ \,}
\def\rtk{\, \right  ] }
\def\rtp{\, \right  ) }
\def\rtb{\, \right \} }

\def\d3x{d^3 x}

\def\cmma{\;\;\; ,}
\def\perd{\;\;\; .}

\def\cm#1{\, {\rm cm^{#1}}}